\documentclass[preprint,12pt]{elsarticle}
\usepackage{amsmath,amsfonts,bm}

\usepackage{array}
\usepackage[caption=false,font=footnotesize,labelfont=rm,textfont=rm]{subfig}
\usepackage{textcomp}
\usepackage{stfloats}
\usepackage{url}
\usepackage{verbatim}
\usepackage{graphicx}

\usepackage{amssymb}
\usepackage{amsthm}
\usepackage[american]{babel}
\usepackage{microtype}
\usepackage{hyperref} 

\usepackage{booktabs}
\usepackage{multirow}
\usepackage{makecell}
\usepackage{xcolor}
\usepackage{pifont}
\usepackage{tabularx}

\usepackage[ruled]{algorithm2e}

\usepackage{threeparttable}

\begin{document}

\begin{frontmatter}

    \title{A capacity renting framework for shared energy storage considering peer-to-peer energy trading of prosumers with privacy protection
    }
 \author[1]{Yingcong Sun}
 \author[1,2]{Laijun Chen\corref{cor}}\ead{chenlaijun@tsinghua.edu.cn}
 \author[3]{Yue Chen}
 \author[1]{Mingrui Tang}
 \author[1,2]{Shengwei Mei}

\cortext[cor]{Corresponding author}

    \affiliation[1]{organization={Department of Electrical Engineering, Tsinghua University},
    		city={Beijing},
    		postcode={100084}, 
    		country={China}}
    \affiliation[2]{organization={College of Energy and Electrical Engineering, Qinghai University},
    		city={Xining},
    		postcode={810016}, 
    		country={China}} 
    \affiliation[3]{organization={Department of Mechanical and Automation Engineering, The Chinese University of Hong Kong},
    		city={Hong Kong},
    		country={China}}

\begin{abstract}
Shared energy storage systems (ESS) present a promising solution to the temporal imbalance between energy generation from renewable distributed generators (DGs) and the power demands of prosumers. However, as DG penetration rates rise, spatial energy imbalances become increasingly significant, necessitating the integration of peer-to-peer (P2P) energy trading within the shared ESS framework. Two key challenges emerge in this context: the absence of effective mechanisms and the greater difficulty for privacy protection due to increased data communication.
This research proposes a capacity renting framework for shared ESS considering P2P energy trading of prosumers. In the proposed framework, prosumers can participate in P2P energy trading and rent capacities from shared ESS. A generalized Nash game is formulated to model the trading process and the competitive interactions among prosumers, and the variational equilibrium of the game is proved to be equivalent to the optimal solution of a quadratic programming (QP) problem.
To address the privacy protection concern, the problem is solved using the alternating direction method of multipliers (ADMM) with the Paillier cryptosystem. Finally, numerical simulations demonstrate the impact of P2P energy trading on the shared ESS framework and validate the effectiveness of the proposed privacy-preserving algorithm.
\end{abstract}


\end{frontmatter}

\section{Introduction}
The proliferation of distributed generators (DGs), especially distributed photovoltaics (PVs) and wind turbines (WTs), has changed electricity production and consumption patterns  \cite{AI2022119968}. An increasing number of consumers have been converted into prosumers with the installation of DGs \cite{grvzanic2022prosumers}. Energy storage systems (ESS) are considered promising solution to mitigate the temporal imbalances between the intermittent generation of DGs and the power demand of prosumers, while also enhancing the economic viability of DGs \cite{choudhury2022review}. However, many prosumers face difficulties in shouldering the high costs and space requirements associated with individual ESS deployment \cite{dai2021utilization}. The concept of shared ESS, which involves centralized energy storage serving multiple prosumers, is receiving attention in numerous countries as it can effectively tackle these challenges with scale effect \cite{wang2024two}.

Currently, energy transaction and capacity allocation are two main ways of energy storage sharing \cite{xiao2022new}. In \cite{ma2024optimal}, the energy transaction framework is employed to enable users to share ESS with VCG mechanism. However, the energy transaction framework cannot directly reflect the prosumers' demand for ESS. The capacity allocation method allows consumers to rent part of the shared ESS for a designated period \cite{zhao2019virtual}, better reflecting their need for regulation capacity. In \cite{chen2022cooperative}, a cooperative game-based approach is applied to allocate shared battery and thermal ESS capacities across various integrated energy systems, with the Nash bargaining method determining the leasing price for capacity. In \cite{jo2020demand}, an energy capacity trading and operation game is proposed to allocate the ESS capacity based on the prosumers' bids. In \cite{xiao2022new}, prosumers rent storage and power capacities separately, further enhancing the flexibility and efficiency of shared energy storage utilization. Collectively, these studies demonstrate that shared ESS can reduce the operational costs for both prosumers and society.

Privacy protection is another important concern in the structure of shared ESS due to the interaction among different stakeholders. Shared ESS can earn more revenue through price discrimination while reducing prosumers' profits if shared ESS can obtain more information from prosumers \cite{lai2021individualized}. In \cite{zheng2022peer}, a centralized solution is used to address capacity allocation for community ESS, which requires users' private information, raising the risk of privacy disclosure. To mitigate such risks, many studies have employed distributed algorithms to avoid the transmission of sensitive parameters between agents. In \cite{jo2020demand}, the NI-function type method is applied to find the variational equilibrium of the proposed game for capacity allocation. In \cite{SUN2024110406}, a hybrid distributed optimization method based on intelligent heuristic algorithms and mathematical programming is proposed to solve the problem of coordinated dispatching of shared ESS, microgrids, and distributed networks. In \cite{LI2022104710}, algorithm of alternative direction multiplier method (ADMM) is applied for benefit allocation among shared ESS and users, where only iterative variable information is exchanged between agents, reducing the risk of privacy leaks. However, even in such cases, private data can sometimes be inferred from iterative data \cite{9316966}. In \cite{zhou2023Coordination}, an example is provided to show how private information is inferred through the iterative data of the dual decomposition based method.

As the penetration of distributed generators (DGs) among prosumers increases, spatial energy imbalances have become a significant challenge. Peer-to-peer (P2P) energy trading has emerged as a cost-effective solution to address this issue \cite{pena2022integration}. In \cite{9618645}, a P2P energy sharing mechanism based on a generalized Nash game model is proposed to increase users' flexibility. In \cite{chen2023region}, the prosumers are aggregated into several regions, and P2P energy transactions are executed among the aggregators. To enhance economic efficiency, it becomes essential to integrate P2P energy trading into the shared ESS framework, thereby addressing both temporal and spatial imbalances in energy generation and demand. This integration leads to more complex interactions. Both energy and capacity are traded, and prosumers are coupled deeply. Conventional distributed algorithms like ADMM are insufficient in this context due to the complex interdependencies \cite{yan2023distributed}, necessitating a new structure to support and describe these interactions.

In the new structure, more information transmission raises more concerns about privacy protection. Distributed optimization alone is no longer sufficient, and techniques such as differential privacy \cite{7525222} and homomorphic encryption \cite{LU2018314} are required. Among them, homomorphic encryption provides greater security and privacy with no artificial noise \cite{10589503}. In \cite{zhang2018admm}, Zhang et al. modify traditional ADMM to incorporate homomorphic encryption for privacy protection. However, this approach is limited in applicability when optimization problems include constraints.

Considering the shortcomings of the existing literature, this research aims to propose a demand-side market mechanism that integrates P2P energy trading into the capacity sharing framework of shared ESS.  The market structure and rules are outlined, and the market participants (prosumers and shared ESS) are modeled. The trading process is formulated as a generalized Nash game among prosumers, which is then transformed into a quadratic programming (QP) problem conditions to solve the equilibrium. To address privacy concerns, a distributed solution algorithm using ADMM with the Paillier cryptosystem is applied. Finally, numerical simulations demonstrate the effectiveness and validity of the proposed mechanism.

The main contributions of the research can be summarized below:

1) A capacity renting framework of shared ESS considering P2P energy trading of prosumers is proposed. In this framework, prosumers can rent capacity from shared ESS and trade energy with other prosumers. A model based on a generalized Nash game is developed to describe the trading process and competition among prosumers. The existence of the variational equilibrium for this game is demonstrated, and the variational equilibrium is proved to be equivalent to the optimal solution of a QP problem. To the best of our knowledge, such existence and equivalence guarantees have not been provided by existing literature. By applying this demand-side mechanism, prosumers' operational costs are reduced.

2) The ADMM algorithm with the Paillier cryptosystem is proposed to solve the equilibrium of the generalized Nash game. The game is transformed into a two-block coupled problem, and the ADMM algorithm with the Paillier cryptosystem is employed to quickly reach equilibrium while safeguarding the privacy of each prosumer. The proposed algorithm can be applied on optimization problems including constraints, and has a convergence speed similar to that of the traditional ADMM algorithm. Through homomorphic encryption, private information remains secure, preventing agents from inferring any private information from communicated information.


\section{System modeling and problem formulation}
\label{System Section}

In this section, we propose the capacity renting framework for shared ESS considering P2P energy trading of prosumers, and develop a deterministic model based on generalized Nash game to describe the trading process.

\subsection{System description}
\begin{figure}
    \centering
    \includegraphics[width = 0.6\linewidth,trim = 220 80 250 140,clip]{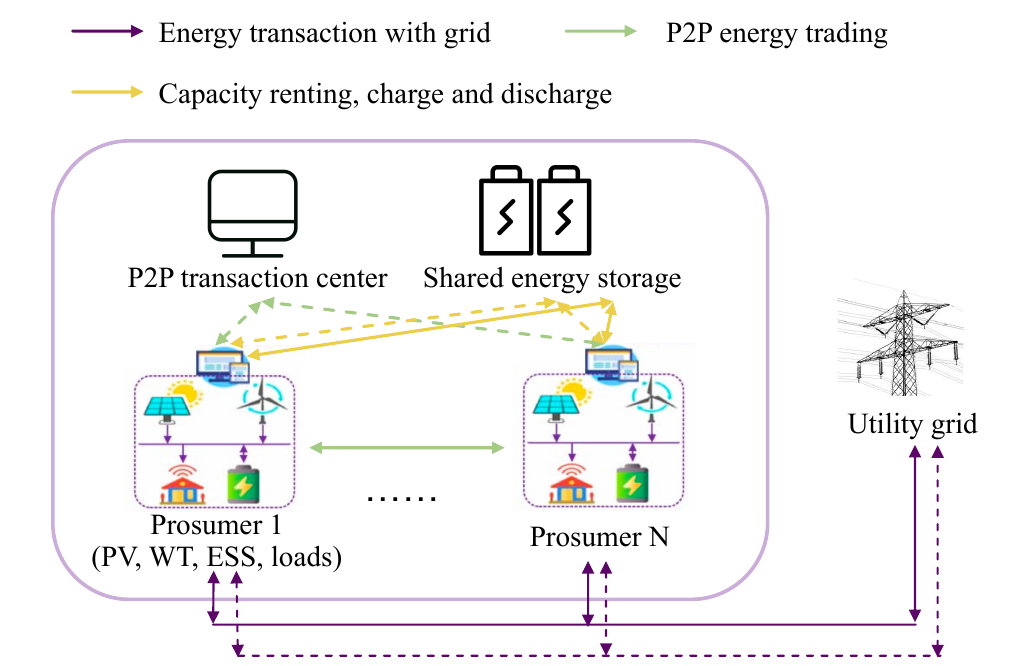}
    \caption{Profiles of energy and capacity transactions. The solid line means energy and capacity transactions, and the dashed line means data communication about energy and capacity transactions.}
    \label{Market profile}
\end{figure}

We consider a local day-ahead market with $N$ prosumers (indexed by $i \in \mathbb{S}$) and shared ESS operated by independent operators. The prosumers are equipped with distributed generators (e.g. PVs and WTs), private ESS and flexible loads. The profiles of energy and capacity transactions are illustrated in fig. \ref{Market profile}. Prosumers can buy electricity from the utility grid or sell electricity to the utility grid. Prosumers can also conduct peer-to-peer energy trading with each other to earn more profits. Shared ESS provides capacity rental services to prosumers, who can rent a portion of the shared ESS and dispatch it to meet their demand. Prosumers are charged based on the amount of capacity rented from the shared ESS. In this framework, prosumers are deeply interconnected through P2P energy trading. Dashed lines in fig. \ref{Market profile} show the communication network among all the agents. A non-profit P2P transaction center manages the transaction process \cite{8864035}. Acting as both information intermediary and supervisory authority, the center ensures that prosumers do not need to communicate directly with each other and oversees the fairness of transactions. In this issue, the impact of uncertainty is relatively small, so deterministic models are used.

\subsection{Models for prosumers' equipment}

Prosumers are equipped with DGs, private ESS, and flexible loads. Suppose that the electricity produced by DGs and the base loads are predicted accurately. For private ESS, the following constraints need to be satisfied:
\begin{subequations}
    \begin{align}
        &S^{i,ES}_{t} = S^{i,ES}_{t-1} + (\eta_{i,ES}P^{i,ES}_{ch,t} - P^{i,ES}_{dis,t}/\eta_{i,ES}) \Delta t
    \label{private energy storage SP constrain}\\
    &0 \leq P^{i,ES}_{ch,t} \leq P^{i,ES}_{max}
    \label{private energy storage discharge bound}\\
    &0 \leq P^{i,ES}_{dis,t} \leq P^{i,ES}_{max}
    \label{private energy storage charge bound}\\
    &SoC_{t}^{i,ES} = S^{i,ES}_{t} / S^{i,ES}_{max}
    \label{SoC definition}\\
    &SoC_{min}^{i,ES} \leq SoC_{t}^{i,ES} \leq SoC_{max}^{i,ES}
    \label{private energy storage SoC bound}
    \end{align}
\end{subequations}
where $P^{i,ES}_{ch,t}, P^{i,ES}_{dis,t}$ are the power charged and discharged of the $i$th prosumer's private ESS, respectively; $S^{i,ES}_{t}, SoC_{t}^{i,ES}$ are the energy stored in and the state of charge of the $i$th prosumer's private ESS at $t$, respectively. $\eta_{i,ES}, P^{i,ES}_{max}, S^{i,ES}_{max}, SoC_{min}^{i,ES}, SoC_{max}^{i,ES}$ are the efficiency, the power capacity, the storage capacity, the lower SoC bound and the upper SoC bound of the $i$th prosumer's private ESS, respectively. 

The usage cost of private ESS is $C_{ES}^{i}$:
\begin{equation}
    C_{ES}^{i} = \sum_{t=1}^{T} \lambda_{u}^{i,ES}(P_{ch,t}^{i,ES} + P_{dis,t}^{i,ES})
\end{equation}
where $\lambda_{u}^{i,ES}$ is the unit power usage price of private ESS.

Suppose part of the prosumers' loads can be shifted to another period, and the following constraints need to be satisfied:
\begin{subequations}
    \begin{align}
    &-\alpha^{i}_{sh} P_{load,t}^{i} <= P^{i}_{sh,t} <= \alpha^{i}_{sh} P_{load,t}^{i}, \quad \forall t
    \label{shiftable load bound}\\
    &\sum_t P^{i}_{sh,t} = 0
    \label{shiftable load sum}
    \end{align}
\end{subequations}
where $P_{load,t}^{i}, P^{i}_{sh,t}$ are the base load and shifted load of prosumer $i$ at $t$; $\alpha^{i}_{sh}$ is the ratio of maximum load that can be shifted.

Let the cost for shifting loads be $C_{sh}^i$:
\begin{equation}
    C_{sh}^i = \sum_{t=1}^{T} \lambda_{sh}^i (P^i_{sh,t}){}^2
\end{equation}
where $\lambda_{sh}^i$ is the price coefficient of shifting loads \cite{he2021new}.

Let $P_{t}^{i}$ be the net loads of prosumer $i$:
\begin{equation}
    P_{t}^{i} = P_{load,t}^{i} + P_{ch,t}^{i,ES} - P_{dis,t}^{i,ES} - P_{DG,t}^{i} + P^{i}_{sh,t}, \quad \forall t
    \label{Prosumer net power}
\end{equation}
where $P_{DG,t}^{i}$ is the power generated from the distributed generator of prosumer $i$.

\subsection{Capacity renting of shared ESS}
The shared ESS allocate power capacity $P^{i,SES}_{max}$ and storage capacity $S^{i,SES}_{max}$ to prosumers. Each prosumer can independently dispatch this capacity for their own use. The constraints on the allocated capacity for prosumers are similar to those applied to private ESS and are outlined below.
\begin{subequations}
    \begin{align}
    &S^{i,SES}_{t} = S^{i,SES}_{t-1} + (\eta_{SES}P^{i,SES}_{ch,t} - P^{i,SES}_{dis,t}/\eta_{SES}) \Delta t
    \label{shared energy storage SP constrain}\\
    &0 \leq P^{i,SES}_{ch,t} \leq P^{i,SES}_{max}
    \label{shared energy storage discharge bound} \\
    &0 \leq P^{i,SES}_{dis,t} \leq P^{i,SES}_{max}
    \label{shared energy storage charge bound}\\
    &SoC_{t}^{i,SES} = S^{i,SES}_{t} / S^{i,SES}_{max}
    \label{SoC definition for SES}\\
    &SoC_{min}^{SES} \leq SoC_{t}^{i,SES} \leq SoC_{max}^{SES}
    \label{shared energy storage SoC bound}\\
    &S^{i,SES}_{T} \geq S^{i,SES}_{0}
    \label{Energy balance for SES}
    \end{align}
\end{subequations}
where $P^{i,SES}_{ch,t}, P^{i,SES}_{dis,t}$ are the charging and discharging power of shared ESS from prosumer $i$, respectively; $S^{i,SES}_{t}, SoC_{t}^{i,SES}$ are the energy belong to prosumer $i$ and the state of charge, respectively. $\eta_{SES}, SoC_{min}^{SES}, SoC_{max}^{SES}$ are the efficiency,  the lower SoC bound and the upper SoC bound of shared ESS, respectively.

The cost of renting and using shared ESS for prosumers $C_{SES}^{i}$ includes three parts: capacity renting cost $C_{SES,r}^{i}$ for renting the capacity of shared ESS, usage cost $C_{SES,u}^{i}$ for shared ESS maintaining, and transmission cost $C_{SES,tr}^i$ to pay for the distribution grid.  
\begin{subequations}
    \begin{align}
        &C_{SES}^{i} = C_{SES,r}^{i} + C_{SES,u}^{i} + C_{SES,tr}^i\\
        &C_{SES,r}^{i} = \lambda^{SES}_{s}S_{max}^{i,SES} + \lambda^{SES}_{p}P_{max}^{i,SES}\\
        &C_{SES,u}^{i} = \sum_{t=1}^{T} \lambda_{u}^{SES}(P_{ch,t}^{i,SES} + P_{dis,t}^{i,SES})\\
        &C_{SES,tr}^i =\sum_{t=1}^{T} \lambda_{tr}^{i,SES} (P_{ch,t}^{i,SES} + P_{dis,t}^{i,SES})
    \end{align}
\end{subequations}
where $\lambda^{SES}_{s}, \lambda^{SES}_{p}$ are the price for unit storage capacity and power capacity, respectively; $\lambda_{u}^{SES}$ is the unit power usage price of shared ESS; $\lambda_{tr}^{i,SES}$ is the unit price for power transmission between prosumer $i$ and shared ESS. 

To reflect the impact of supply and demand on prices, the capacity renting price is determined by the following equations:
\begin{subequations}
    \begin{align}
        &\lambda^{SES}_{s} = a_s  + b_s\sum_{i \in \mathbb{S}} S_{max}^{i,SES}\\
        &\lambda^{SES}_{p} = a_p  + b_p\sum_{i \in \mathbb{S}} P_{max}^{i,SES}
    \end{align}
\end{subequations}
where $a_s, b_s, a_p, b_p$ are the price coefficient of shared ESS, which is predetermined before renting. When there is a tight supply of capacity of shared ESS, the price will be higher, and vice versa. The equations also show that other prosumers' decisions can affect the cost of renting capacities directly. 

Let the unit power transmission price between $i$ and $j$ be $\lambda_{tr}^{i,j}$:
\begin{equation}
    \lambda_{tr}^{i,j} = \theta_1 |x^{i,j}| + \theta_0
\end{equation}
where $x^{i,j}$ is the impedance between prosumer $i$ and $j$. $\theta_1, \theta_0$ are constant numbers \cite{ullah2021peer}.

The summation of the power capacity and storage capacity allocated to prosumers cannot exceed the maximum capacity of shared ESS:
\begin{subequations}
    \begin{align}
    &\sum_{i \in \mathbb{S}} P^{i,SES}_{max} \leq P^{SES}_{max}
    \label{Shared Energy Storage Power Capacity constrains}\\
    &\sum_{i \in \mathbb{S}} S^{i,SES}_{max} \leq S^{SES}_{max}
    \label{Shared Energy Storage Capacity Capacity constrains}
    \end{align}
\end{subequations}
where $P^{SES}_{max}, S^{SES}_{max}$ are the power capacity and storage capacity of shared ESS, respectively.

\subsection{Transactions in wholesale market and P2P market}

The prosumers can buy electricity from or sell electricity to the utility grids, and the following constraints need to be satisfied:
\begin{subequations}
    \begin{align}
    &P^i_{sell,t} \geq 0
    \label{Power sold bound}\\
    &P^i_{buy,t} \geq 0
    \label{Power bought bound}
    \end{align}
\end{subequations}
where $P^i_{sell,t}, P^i_{buy,t}$ are the $i$th prosumer's power sold to and bought from the utility grid, respectively.

Let $C_{g}^{i}$ be the cost of electricity interaction with the utility grid:
\begin{equation}
    C_{g}^{i} = \sum_{t=1}^{T} (\lambda^{buy}_{t}P_{buy,t}^{i} - \lambda^{sell}_{t}P_{sell,t}^{i})
\end{equation}
where $\lambda^{buy}_{t}, \lambda^{sell}_{t}$ are the buying and selling price of electricity, respectively. The selling price is usually much lower than the buying price, and therefore the prosumers will not buy and sell electricity at the same time.

The prosumers can also trade electricity with other prosumers with a price between buying price and selling price. Let $P^{i,j}_t$ be the power prosumer $i$ bought from prosumer $j$. Therefore, The sum of every two prosumers' electricity transactions is 0:
\begin{equation}
    P^{i,j}_t + P^{j,i}_t = 0
    \label{P2P power equality}
\end{equation}

The cost of P2P trading $C_{P2P}^{i}$ can be calculated as below:
\begin{equation}
    C_{P2P}^{i} = \sum_{t=1}^{T}\sum_{j \in \mathbb{S}}\lambda^{P2P}_{t}P_{t}^{i,j}
\end{equation}
where $\lambda^{P2P}_{t}$ is the P2P trading price. For simplicity, the price is set at:
\begin{equation}
    \lambda^{P2P}_{t} = (\lambda^{buy}_{t} + \lambda^{sell}_{t}) / 2
\end{equation}

The prosumer who buys electricity from P2P energy trading should pay the transmission fee $C_{tr}^i$.
\begin{equation}
    C_{tr}^i =\sum_{t=1}^{T}\sum_{j \in \mathbb{S}} \lambda_{tr}^{i,j} max\{0,P_t^{i,j}\}
    \label{Transmission fee}
\end{equation}

Eqs. (\ref{Transmission fee}) is neither affine nor quadratic, and it can be linearized via introducing assistance variables $P^{i,j,+}_{t}$. Turn eqs. (\ref{Transmission fee}) into:
\begin{equation}
    C_{tr}^i =\sum_{t=1}^{T}\sum_{j \in \mathbb{S}} \lambda_{tr}^{i,j} P_t^{i,j,+}
\end{equation}
where
\begin{equation}
    P_t^{i,j,+} \geq P_t^{i,j}
\end{equation}
\begin{equation}
    P_t^{i,j,+} \geq 0
\end{equation}

Through buying and selling electricity with the utility grid and other prosumers, the prosumers should achieve power balance:
\begin{equation}
    P_t^i = \sum_{j \in \mathbb{S}} P^{i,j}_t + P^i_{buy,t} + P^{i,SES}_{dis,t} - P^i_{sell,t} - P^{i,SES}_{ch,t} 
    \label{Prosumer balance}
\end{equation}

\subsection{Generalized Nash Game Model}

Let $\bm{x^i} = [\bm{x^i_0},\bm{x^i_1}]$ be the decision variables of prosumer $i$, and let $\bm{x} = [\bm{x^1},...,\bm{x^n}]$. Let $\bm{x^{-i}}$ be the decision variables of prosumers except prosumer $i$. 
\begin{align*}
    \bm{x^i_0} =& \big{[} P_{ch,t}^{i,ES}, P_{dis,t}^{i,ES},  P_{ch,t}^{i,SES}, P_{dis,t}^{i,SES}, \notag \\
    &  P_{buy,t}^{i}, P_{sell,t}^i,P_{t}^{i,j},P_t^{i,j,+},P_{sh,t}^{i} \big{]} \notag \\
    \bm{x^i_1} =& \big{[}P_{max}^{i,SES}, S_{max}^{i,SES} \big{]}
\end{align*}
where $\bm{x^i_0}$ are the decision variables of prosumer $i$ that do not directly affect the costs of other prosumers, and $\bm{x^i_1}$ are the decision variables of prosumer $i$ that directly impact the costs of other prosumers. Let $C^{i}_0(\bm{x^i_0})$ be the cost only determined by the $i$th prosumers' own decisions and $C^{i}_1(\bm{x^i_1}, \bm{x^{-i}_1})$ be the cost determined by all prosumers' decisions.

\begin{equation*}
   C^{i}_0(\bm{x^i_0}) = C_{g}^{i} + C_{P2P}^{i} + C_{SES,u}^{i} + C_{SES,tr}^i + C_{ES}^{i} + C_{tr}^{i} + C_{sh}^i   
\end{equation*}
\begin{align*}
    &C^{i}_1(\bm{x^i_1}, \bm{x^{-i}_1}) = C_{SES}^{r} =\notag \\ 
    & (a_s  + b_s\sum_{j \in \mathbb{S}} S_{max}^{j,SES}) S_{max}^{i,SES} 
    + (a_p  + b_p\sum_{j \in \mathbb{S}} P_{max}^{j,SES}) P_{max}^{i,SES} 
\end{align*}

For each prosumer, his objective is to minimize the total cost. Let $G = \langle \mathbb{S}, \bm{X}, \bm{C} \rangle$ be the generalized Nash game, where $\mathbb{S}$ is the player set, $\bm{X} = \{\bm{X^i}, {i \in \mathbb{S}}\}$ is the strategy set of each prosumer, and $\bm{C} = \{C^i,{i \in \mathbb{S}}\}$ is the utility function of each prosumer. Prosumers need to solve the problem (\ref{Initial problem}). 

\begin{align}
    \min_{\bm{x^i}} \quad & C^{i}(\bm{x^i}, \bm{x^{-i}}) = C^{i}_0(\bm{x^i_0}) + C^{i}_1(\bm{x^i_1}, \bm{x^{-i}_1})\notag \\
    s.t. \quad &\bm{x^i} \in \bm{X^i} = \bm{X^i_0} \cap \bm{X^i_1}(\bm{x^{-i}})
    \label{Initial problem}
\end{align}
where $\bm{X^i_0}$ shows the decision space of prosumer $i$ not coupled with other prosumers', and $\bm{X^i_1}(\bm{x^{-i}})$ shows the coupled decision space.
\begin{equation*}
    \bm{X^i_0} = \{\bm{x^i}: (1),(3),(5),(6),(11),(20)\}
\end{equation*}
\begin{equation*}
    \bm{X^i_1}(\bm{x^{-i}}) = \{\bm{x^i}: (10),(13)\}
\end{equation*}

The generalized Nash game model captures the competition among prosumers in the P2P market for shared ESS capacity and electricity, as each prosumer seeks to minimize their individual costs. Prosumers have full control over their private energy storage systems (ESS), the shared ESS they have rented, and their shiftable loads.

\section{Solution algorithm for Generalized Nash equilibrium}
\label{Solution Section}

In this section, we propose a distributed algorithm to solve the variational equilibrium of the game with privacy protection.

\subsection{Equivalent transformation to a 2-block QP problem}

\textbf{Proposition 1}: The variational equilibrium of the generalized Nash game (\ref{Initial problem}) exists if and only if problem (\ref{Centralized problem}) has a solution. If the variational equilibrium of (\ref{Initial problem}) exists, the variational equilibrium is equivalent to the optimal solution of (\ref{Centralized problem}).
\begin{align}
    \min_{\bm{x}}\quad & \sum_{i \in \mathbb{S}} C_0^i(\bm{x_0^i}) + \frac{1}{2} b_s (\sum_{i \in \mathbb{S}} S_{max}^{i,SES})^2 + a_s(\sum_{i \in \mathbb{S}} S_{max}^{i,SES}) \notag \\
    &+ \frac{1}{2}b_s(\sum_{i \in \mathbb{S}} {S_{max}^{i,SES}}^2)  + \frac{1}{2} b_p (\sum_{i \in \mathbb{S}} P_{max}^{i,SES})^2 \notag \\
    &+ a_p(\sum_{i \in \mathbb{S}} P_{max}^{i,SES}) + \frac{1}{2}b_p(\sum_{i \in \mathbb{S}} {P_{max}^{i,SES}}^2) \notag \\
    s.t. \quad& \bm{x^i} \in \bm{X^i} \quad \forall i \in \mathbb{S}
    \label{Centralized problem}
\end{align}

Proof: see Appendix A.

Problem (\ref{Centralized problem}) provides a centralized way to solve the GNE, which needs to collect each participant's private information. For privacy protection, a decentralized method based on ADMM is proposed. First, the problem (\ref{Centralized problem}) must be reformulated as a 2-block problem, making it suitable for ADMM \cite{yan2023distributed}.

Introduce assistant variables in the transaction center: $\tilde{P}^{i,j}_{t}$. Use the following equations to replace (\ref{P2P power equality}):
\begin{align}
    &\tilde{P}^{i,j}_{t} = {P}^{i,j}_{t}
    \label{P2P assi equality} \\
    &\tilde{P}^{i,j}_{t} + \tilde{P}^{j,i}_{t} = 0
    \label{P2P center constrains}
\end{align}

Introduce assistant variables in shared ESS: $\tilde{P}^{i,SES}_{max}, \tilde{S}^{i,SES}_{max}$, and use the following equations to replace eqs. (\ref{Shared Energy Storage Power Capacity constrains}) and (\ref{Shared Energy Storage Capacity Capacity constrains}).
\begin{align}
    &\tilde{P}^{i,SES}_{max} = {P}^{i,SES}_{max}
    \label{SES-P assi equality} \\
    &\tilde{S}^{i,SES}_{max} = {S}^{i,SES}_{max}
    \label{SES-S assi equality} \\
    &\sum_{i \in \mathbb{S}} \tilde{P}^{i,SES}_{max} \leq P^{SES}_{max}
    \label{SES-P constrain} \\
    &\sum_{i \in \mathbb{S}} \tilde{S}^{i,SES}_{max} \leq S^{SES}_{max}
    \label{SES-S constrain}
\end{align}

Rewrite problem (\ref{Centralized problem}) as problem (\ref{two-block}). In problem (\ref{two-block}), only the decision variables of each prosumer and shared ESS, each prosumer and transaction center are coupled, and the objective is a 2-block structure. Therefore, algorithm of ADMM can be applied. The standard process of the algorithm of ADMM can be seen in Appendix B. 
\begin{align}
    \min & \; \sum_{i \in \mathbb{S}} C_0^i(\bm{x_0^i}) +
    \frac{1}{2} b_s (\sum_{i \in \mathbb{S}} \tilde{S}_{max}^{i,SES})^2 + a_s(\sum_{i \in \mathbb{S}} \tilde{S}_{max}^{i,SES}) \notag \\
    &+ \frac{1}{2}b_s(\sum_{i \in \mathbb{S}} \tilde{S}{{}_{max}^{i,SES}}^2)
    +\frac{1}{2} b_p (\sum_{i \in \mathbb{S}} \tilde{P}_{max}^{i,SES})^2 \notag \\
    &+ a_p(\sum_{i \in \mathbb{S}} \tilde{P}_{max}^{i,SES}) + \frac{1}{2}b_p(\sum_{i \in \mathbb{S}} \tilde{P}{{}_{max}^{i,SES}}^2) \notag \\
    s.t. &\; \bm{x^i} \in \bm{X_0^i} \quad \forall i \in \mathbb{S}, \quad (23)-(28) 
    \label{two-block}
\end{align}

\subsection{Algorithm of ADMM with Paillier cryptosystem}
In the process of standard ADMM, although prosumers are not required to directly share their private data (e.g. $\lambda_{sh}^i$) with others, relative information can still be obtained through the regular sequence intermediate states data (e.g. $P_{max}^{i,SES}(k)$ transferred to shared ESS), as demonstrated in Appendix B. Therefore, the intermediate states of each agent should also be regarded as privacy and be protected \cite{zhang2018admm}. The algorithm of ADMM with Paillier cryptosystem is applied, where the intermediate states of each agent are transferred in encrypted form to protect privacy. The algorithm is detailed in Algorithm \ref{Encrypted ADMM}. The algorithm is modified based on the algorithm of standard ADMM, and the derivation is provided in Appendix C.

For each prosumer, they need to update variables $\bm{x^{i}}$ by (\ref{Encry prosumer update}).
\begin{align}
    &\bm{x^{i}}(k+1) = \arg \min_{\bm{x^i} \in \bm{X^i}} \quad C^{i}_0(\bm{x^i}) 
     \notag \\
    &+ \frac{\beta_P^i}{2}\parallel P_{max}^{i,SES} - P_{max}^{i,SES}(k)\parallel_2^2 + (\alpha_P^i(k)-\mu_P^i(k))P_{max}^{i,SES} \notag \\
    &+ \frac{\beta_S^i}{2}\parallel S_{max}^{i,SES} - S_{max}^{i,SES}(k)\parallel_2^2 + (\alpha_S^i(k)-\mu_S^i(k))S_{max}^{i,SES} \notag \\
    &+ \sum_{j \in \mathbb{S},t}(\frac{\beta_{P2P}^i}{2}\parallel {P}^{i,j}_{t} - {P}^{i,j}_{t}(k)\parallel_2^2 + (\alpha_t^{i,j}(k)-\mu_{t}^{i,j}(k)){P}^{i,j}_{t})
    \label{Encry prosumer update}
\end{align}
where $\beta_P^i, \beta_S^i, \beta_{P2P}^i$ are constant numbers and private for prosumer $i$. ${\mu}_P^i, {\mu}_S^i$ and $ {\mu}^{i,j}_t$ are the dual variables of (\ref{SES-P assi equality}), (\ref{SES-S assi equality}), and (\ref{P2P assi equality}), respectively. $\alpha_P^i, \alpha_S^i, \alpha_{t}^{i,j}$ are calculated through (\ref{alpha compute}) in encrypted forms with Paillier cryptosystem. The computation process needs to communicate with shared ESS and the P2P transaction center and is shown in algorithm \ref{Encrypted calculation}. The Paillier cryptosystem \cite{paillier1999public} is described in Appendix D.
\begin{align}
    &\alpha_P^i(k) = \tau_P^i(k)\tilde{\tau}_P^i(k)(P_{max}^{i,SES}(k) - \tilde{P}_{max}^{i,SES}(k)) \notag \\
    &\alpha_S^i(k) = \tau_S^i(k)\tilde{\tau}_S^i(k)(S_{max}^{i,SES}(k) - \tilde{S}_{max}^{i,SES}(k)) \notag \\
    &\alpha_{t}^{i,j}(k) = \tau_{P2P}^i(k)\tilde{\tau}_{P2P}^i(k)(P_t^{i,j}(k) - \tilde{P}_t^{i,j}(k))
    \label{alpha compute}
\end{align}
where $\tau_P^i, \tau_S^i, \tau_{P2P}^i$ are random numbers generated by prosumer $i$, $\tilde{\tau}_P^i, \tilde{\tau}_S^i$ are random numbers generated by shared ESS, $\tilde{\tau}_{P2P}^i$ are random numbers generated by P2P transaction center.

The shared ESS updates the variables $\bm{\tilde{P}^{SES}}(k+1) = \{\tilde{P}{{}_{max}^{i,SES}}, \forall i \in \mathbb{S} \}, \bm{\tilde{S}^{SES}}(k+1) = \{\tilde{S}{{}_{max}^{i,SES}}, \forall i \in \mathbb{S} \}$ by (\ref{Encry SES-P update}), (\ref{Encry SES-S update})
\begin{align}
    &\bm{\tilde{P}^{SES}}(k+1) = \arg \min_{\bm{\tilde{P}^{SES}}} \frac{1}{2} b_p (\sum_{i \in \mathbb{S}} \tilde{P}_{max}^{i,SES})^2 \notag \\
    &+ a_p(\sum_{i \in \mathbb{S}} \tilde{P}_{max}^{i,SES}) + \frac{1}{2}b_p(\sum_{i \in \mathbb{S}} \tilde{P}{{}_{max}^{i,SES}}^2) \notag \\
    &+ \sum_{i \in \mathbb{S}} \frac{\tilde{\beta}_P^i}{2}\parallel \tilde{P}_{max}^{i,SES} - \tilde{P}_{max}^{i,SES}(k)\parallel_2^2 +(\tilde{\alpha}_P^i(k)+\mu_P^i(k))\tilde{P}_{max}^{i,SES}\notag \\
    &s.t. \quad \text{eqs}. (\ref{SES-P constrain})
    \label{Encry SES-P update}
\end{align}
\begin{align}
    &\bm{\tilde{S}^{SES}}(k+1) = \arg \min_{\bm{\tilde{S}^{SES}}} \frac{1}{2} b_s (\sum_{i \in \mathbb{S}} \tilde{S}_{max}^{i,SES})^2 \notag \\
    &+ a_s(\sum_{i \in \mathbb{S}} \tilde{S}_{max}^{i,SES}) + \frac{1}{2}b_s(\sum_{i \in \mathbb{S}} \tilde{S}{{}_{max}^{i,SES}}^2) \notag \\
    &+ \sum_{i \in \mathbb{S}} \frac{\tilde{\beta}_S^i}{2}\parallel \tilde{S}_{max}^{i,SES} - \tilde{S}_{max}^{i,SES}(k)\parallel_2^2 +(\tilde{\alpha}_S^i(k)+\mu_P^i(k))\tilde{S}_{max}^{i,SES}\notag \\
    &s.t. \quad \text{eqs}. (\ref{SES-S constrain})
    \label{Encry SES-S update}
\end{align}
where $\tilde{\beta}_P^i, \tilde{\beta}_S^i$ are constant numbers. $\tilde{\alpha}_P^i, \tilde{\alpha}_S^i$ are calculated through eqs (\ref{SES alpha compute}) and algorithm \ref{Encrypted calculation}.
\begin{align}
    & \tilde{\alpha}_P^i(k) = \tilde{\tau}_P^i(k){\tau}_P^i(k)(\tilde{P}_{max}^{i,SES}(k) - {P}_{max}^{i,SES}(k+1)) \notag \\
    &\tilde{\alpha}_S^i(k) = \tilde{\tau}_S^i(k){\tau}_S^i(k)(\tilde{S}_{max}^{i,SES}(k) - {S}_{max}^{i,SES}(k+1))
    \label{SES alpha compute}
\end{align}

P2P transaction center updates the variables $\bm{\tilde{P}_{P2P}} = \{\tilde{P}^{i,j}_{t}, \forall i,j \in \mathbb{S}, t\}$ by (\ref{Encry P2P update}).
\begin{align}
    &\bm{\tilde{P}_{P2P}}(k+1) = \arg \min_{\bm{\tilde{P}_{P2P}}} \notag \\
    &\sum_{i,j \in \mathbb{S}}(\frac{\tilde{\beta}_{P2P}^i}{2}\parallel \tilde{P}^{i,j}_{t} - \tilde{P}^{i,j}_{t}(k)\parallel_2^2 + (\tilde{\alpha}_t^{i,j}(k)+\mu_{t}^{i,j}(k))\tilde{P}^{i,j}_{t}) \notag \\
    &s.t. \quad  \text{eqs}. (\ref{P2P center constrains})
    \label{Encry P2P update}
\end{align}
where $\tilde{\beta}_{P2P}^i$ are constant numbers and private for P2P transaction center. 
\begin{equation}
    \tilde{\alpha}_{t}^{i,j}(k) = \tilde{\tau}_{P2P}^i(k){\tau}_{P2P}^i(k)(\tilde{P}_t^{i,j}(k) - {P}_t^{i,j}(k+1))
    \label{P2P alpha compute}
\end{equation}

Then, the dual variables are updated by eqs. (\ref{Encry dual update}) with algorithm \ref{Encrypted calculation}. The dual variables are consistent among prosumers, shared ESS, and P2P transaction centers.
\begin{align}
    &\mu_P^{i}(k+1) = {\mu}_P^i(k) + (1+e^{-rk})\tau_P^i(k)\tilde{\tau}_P^i(k)(\tilde{P}_{max}^{i,SES}(k+1) - {P}_{max}^{i,SES}(k+1)) \notag \\
    &\mu_S^{i}(k+1) = {\mu}_S^i(k) + (1+e^{-rk})\tau_S^i(k)\tilde{\tau}_S^i(k)(\tilde{S}^{i,SES}_{max}(k+1) - {S}^{i,SES}_{max}(k+1)) \notag \\
    &\mu^{i,j}_t(k+1) = {\mu}^{i,j}_t(k) + (1+e^{-rk})\tau_{P2P}^i(k)\tilde{\tau}_{P2P}^i(k) (\tilde{P}^{i,j}_{t}(k+1) - {P}^{i,j}_{t}(k+1)) \notag \\
    \label{Encry dual update}
\end{align}
where $r$ is a constant number for acceleration. The convergence is ensured as a quadratic proximal term is introduced in the modified algorithm \cite{deng2017parallel}.

The algorithm can achieve privacy protection in two aspects. In algorithm \ref{Encrypted calculation}, each agent can only obtain information about the product of the intermediate variable and a random number from other agents, thus protecting the information of the intermediate variable. In the process of iteration, $\beta$ is private information, which also adds difficulty to the inference of the objective function.

\begin{algorithm}[!t]
Initialize: 
$\tilde{P}^{i,SES}_{max}(1)=0$,
$\tilde{S}^{i,SES}_{max}(1)=0$,
$\bm{\tilde{P}_{P2P}} = \bm{0}$,
${\mu}_P^i(1)=0$, ${\mu}_S^i(1)=0$, ${\mu}^{i,j}_t(1)=0$, $\beta > 0$,  $\epsilon$

\For{$k=1,2,3...$}{

Shared ESS generates random number $\tilde{\tau}_P^i(k), \tilde{\tau}_S^i(k)$

P2P transaction center generates random number $\tilde{\tau}_{P2P}^i(k)$

    \For{prosumer $i \in \mathbb{S}$}{
    generate random number ${\tau}_P^i(k), {\tau}_S^i(k), {\tau}_{P2P}^i(k)$

    compute $\alpha_P^i(k), \alpha_S^i(k), \alpha_t^{i,j}(k)$ by (\ref{alpha compute}) and algorithm \ref{Encrypted calculation}.

    update $\bm{x^i}(k+1)$ by problem (\ref{Encry prosumer update}).
}

Shared ESS computes $\tilde{\alpha}_P^i(k), \tilde{\alpha}_S^i(k)$ by (\ref{SES alpha compute}) and algorithm \ref{Encrypted calculation}. Then update $\bm{\tilde{P}^{SES}}(k+1),\bm{\tilde{S}^{SES}}(k+1)$ by (\ref{Encry SES-P update}),(\ref{Encry SES-S update}).

P2P transaction center computes $\tilde{\alpha}_t^{i,j}(k)$ by (\ref{P2P alpha compute}) and algorithm \ref{Encrypted calculation}. Then update $\bm{\tilde{P}_{P2P}}(k+1)$ by (\ref{Encry P2P update}).

Each agent update dual variables by (\ref{Encry dual update}) and algorithm \ref{Encrypted calculation}.

Each agent computes the gap: Euclidean distances before and after the update of the dual variable. If all the gaps are less than $\epsilon$, then break.

}

\caption{Algorithm of ADMM with Paillier cryptosystem}
\label{Encrypted ADMM}
\end{algorithm}

\begin{algorithm}[!t]
\KwData{$\tau_1,\theta_1$ private for agent 1, $\tau_2,\theta_2$ private for agent 2}
\KwResult{Agent 1 computes $\alpha_1 = \tau_1\tau_2(\theta_1-\theta_2)$ without knowing the value of $\tau_2,\theta_2$}

Agent 1 generates Paillier key: public key ($\Gamma_1, \Omega_1$), private key ($\zeta_1, \sigma_1$)

Agent 1 computes the ciphertext of $\theta_1$: $E(\theta_1)$, and then transfers $E(\theta_1),\Gamma_1, \Omega_1$ to agent 2. 

Agent 2 computes the ciphertext of $\tau_2(\theta_1-\theta_2)$: $E(\tau_2(\theta_1-\theta_2)) = E(\theta_1)^{\tau_2}E(-\theta_2)^{\tau_2} \; mod \; \Gamma_1^2$, and then transfers the result to agent 1. Only the public key from agent 1 is used.

Agent 1 computes $E(\tau_1\tau_2(\theta_1-\theta_2))$, and decrypt with ($\zeta_1, \sigma_1$) to obtain $\alpha_1$

\caption{Encrypted calculation process of $\alpha$}
\label{Encrypted calculation}
\end{algorithm}

\section{Case study}
\label{Case Section}

\subsection{Test system configuration}

\begin{figure}[!t]
    \centering
    \includegraphics[trim = 2 2 2 2,clip]{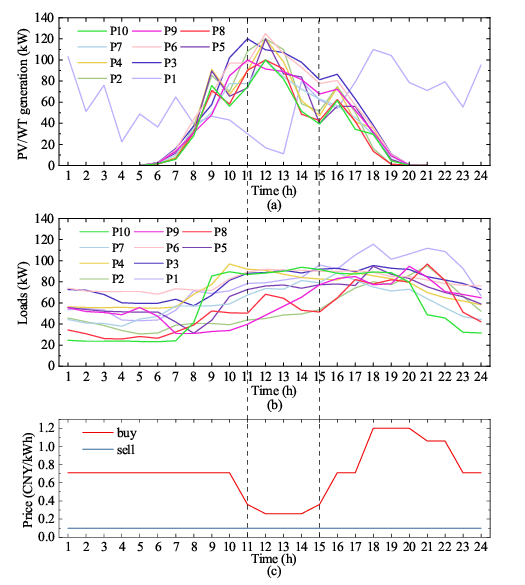}
    \caption{(a) PV/WT generation of prosumers. Only prosumer 1 is equipped with WTs, and others are equipped with PVs. (b) Loads of prosumers (c) The electricity price}
    \label{generation and load}
\end{figure}

The test system is shown in fig. \ref{Market profile}. The participants in the market include 10 prosumers and a 150kW/450kWh shared ESS. Prosumer 1 is equipped with WTs and other prosumers are equipped with PVs. The power generation and load profiles for each prosumer are shown in Fig. \ref{generation and load}. The price coefficients of shiftable loads $\lambda_{sh}^{i}$ range from $0.03$ to $0.05 \text{CNY}/\text{kWh}^2$, and the ratio of shiftable loads is set at 10\%. Prosumer 2 and prosumer 3 are equipped with 10kW/20kWh, and 20kW/40kWh private ESS, respectively. The efficiency of private ESS is 95\%, and the maximum and minimum SoC of private ESS are 10\% and 90\%, respectively. The price of private ESS usage is 0.11 CNY/kW·h. The efficiency, maximum and minimum SoC for the shared ESS are 95\%, 90\%, and 10\%, respectively. The price for power capacity is fixed at 0.3 CNY/kW·day. The price for storage capacity is set at 0.22 CNY/kWh·day when there is no demand, and 0.4 CNY/kWh·day when the demand exceeds supply. The usage price of shared ESS is 0.08 CNY/kW·h. The electricity price is depicted in Fig. \ref{generation and load} \cite{ShandongPrice}, and it aligns with regional supply and demand conditions.


The above models and methods are implemented using Matlab R2022a with Gurobi Optimizer version 9.5.2, on a 2GHz Intel Core i5 CPU with 16GB RAM.

Four cases are carried out in the case study to evaluate the effect of P2P energy trading and shared ESS.

\textbf{Case 1}: P2P energy trading is prohibited, and no shared ESS is available. In this case, each prosumer interacts solely with the utility grid.

\textbf{Case 2}: Prosumers are permitted to engage in P2P energy trading, but no shared ESS is available.

\textbf{Case 3}: P2P energy trading is prohibited, but a 150 kW/500 kWh shared ESS is available for rental.

\textbf{Case 4}: Prosumers are permitted to engage in P2P energy trading and rent the 150 kW/500 kWh shared ESS, in accordance with the model proposed in this study.

\subsection{Benefit of the mechanism}
Table. \ref{Cost participants} shows the revenue of shared ESS and the cost of each prosumer in the three cases. From the perspective of total cost in the market, it can be found that each participant in the market can benefit from shared ESS (compare case 1 with case 3, and case 2 with case 4), and each prosumer can benefit from P2P energy trading as well (compare case 1 with case 2, and case 3 with case 4). From the comparison of case 1,3 and 4, the following conclusions can be drawn. Shared ESS can reduce the total cost by about 5\% (327 CNY/day), and the addition of P2P energy trading further lowers the total cost by 173 CNY/day. This indicates that P2P transactions enhance social welfare beyond the benefits provided by the shared ESS framework. When P2P energy trading is prohibited, the shared ESS can generate up to 245 CNY/day, representing 75\% of the total social benefits of the shared ESS, while the remaining 25\% of benefits are concentrated among a few prosumers (e.g., prosumers 1, 2, and 9). This concentration is attributed to the monopolistic position of the shared ESS and the generation and load characteristics of the prosumers. Allowing P2P energy trading results in only a slight decrease of 7 CNY/day in the revenue of the shared ESS, while significantly reducing electricity costs for prosumers, particularly for prosumer 9 in the case study (The reasons will be analyzed later). Therefore, P2P energy trading provides substantial social benefits, albeit with a minor reduction in the revenue of the shared ESS.
\begin{table}[!t]
    \centering
    \caption{Cost/revenue of each participant in the market}
    \label{Cost participants}
    \begin{threeparttable}
    \begin{tabularx}{\linewidth}{p{2.5cm}>{\centering\arraybackslash}X>{\centering\arraybackslash}X>{\centering\arraybackslash}X>{\centering\arraybackslash}X}
        \toprule
           Participants \tnote{*} & Case 1 & Case 2 & Case 3 & Case 4\\
        \midrule
            Prosumer 1    & 297  & 243  & 271    & 252    \\
            Prosumer 2    & 678  & 650  & 662    & 637    \\
            Prosumer 3    & 870  & 861  & 866    & 857    \\
            Prosumer 4    & 824  & 813  & 822    & 810    \\
            Prosumer 5    & 778  & 771  & 775    & 764    \\
            Prosumer 6    & 904  & 901  & 901    & 897    \\
            Prosumer 7    & 679  & 673  & 675    & 670    \\
            Prosumer 8    & 677  & 670  & 670    & 662    \\
            Prosumer 9    & 737  & 663  & 722    & 651    \\
            Prosumer 10   & 648  & 632  & 648    & 632    \\
            SES \tnote{**} & -   & -    & 245    & 238    \\
        \midrule
            Total         & 7094 & 6877  & 6767   & 6594  \\
        \bottomrule
    \end{tabularx}

    \begin{tablenotes}
        \footnotesize
        \item[*] This table shows the cost of prosumers, revenue of SES, and the total cost of all participants. The unit is CNY.
        \item[**] The data shows the revenue of SES.
    \end{tablenotes}
    \end{threeparttable}
\end{table}


\subsection{Capacity renting of shared ESS in different cases}
\begin{figure}
    \centering
    \includegraphics[trim = 2 2 2 6,clip]{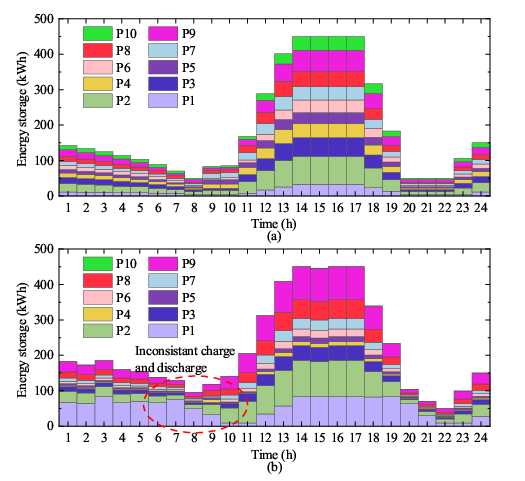}
    \caption{Energy stored in the shared ESS of each prosumer (a) case 4: with P2P (b) case 3: without P2P}
    \label{SOC comparison}
\end{figure}

To show the effect of P2P energy trading on capacity leasing and charge-discharge behavior of prosumers, fig. \ref{SOC comparison} compares the energy stored in the shared ESS of each prosumer in case 4 and case 3. The maximum energy stored represents the leased storage capacity of each prosumer (The energy stored between 14:00-17:00 for most prosumers). P2P energy trading affects the capacity rented by each prosumer: Prosumer 1, 2, 8, and 9 rent less shared ESS when P2P trading is allowed, whereas other prosumers increase their rental of shared ESS. This observation aligns with the benefits of P2P trading depicted in Table \ref{Cost participants}. When P2P energy trading is forbidden, prosumer 1,2,9 rent more capacities than others as they benefit more from renting shared ESS. If P2P energy trading is allowed, P2P transactions have the potential to substitute shared ESS demands. These prosumers can generate earnings from P2P trading and reduce costs by renting fewer shared ESS capacities. For other prosumers, P2P energy trading benefits them by allowing them to rent more shared ESS in the competition, thereby resulting in only marginal cost reductions. 

Additionally, P2P energy trading enhances the consistency of charge and discharge behavior. In fig \ref{SOC comparison} (b), prosumer 1 discharges between 6:00 - 10:00 while others charge, which wastes energy. P2P energy trading makes them charge and discharge simultaneously, showing the efficiency promotion of shared ESS.

\subsection{Operation results of prosumers}

\begin{figure}
    \centering
    \includegraphics[trim = 2 2 2 6,clip]{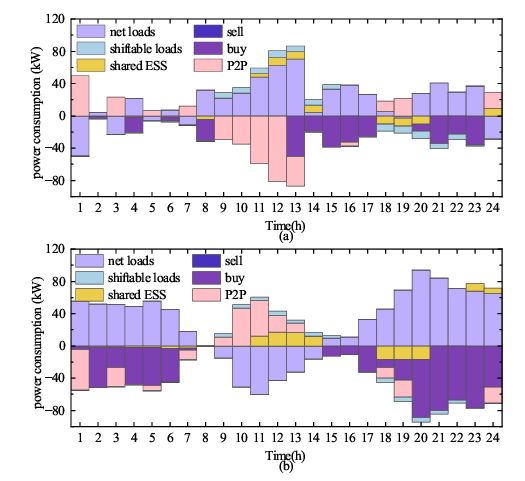}
    \caption{Profiles of energy supply and consumption of (a) prosumer 1 (with WTs)  (b) prosumer 9 (with PVs)}
    \label{Energy profile P1}
\end{figure}

Fig. \ref{Energy profile P1} shows the electricity supply and consumption profiles of prosumer 1 and prosumer 9 in case 4. In the figures, positive values indicate power consumption and negative values indicate power supply. Prosumer 1, who is equipped with WTs, exhibits low net loads between 17:00 and 07:00, a period when electricity prices are relatively high. This low net load is the primary reason for Prosumer 1’s lowest electricity cost. Prosumer 1 can sell electricity at night and buy electricity at noon through P2P energy trading, as other prosumers with PV need energy at night and have excess energy generation at noon. In the shared ESS competition, Prosumer 1 rents a relatively small capacity because he can address most of the imbalance problems through P2P transactions. Prosumer 1 charges the shared ESS at noon by purchasing electricity at a lower price through P2P trading, and discharges the ESS between 18:00 and 20:00 to sell electricity and earn profits. This strategy also reduces the amount of electricity that needs to be purchased at higher prices.

Prosumer 9, a typical PV-equipped prosumer, experiences high net loads between 17:00 and 07:00, resulting in significant electricity purchase costs. Due to lower P2P transaction costs, Prosumer 9 engages more frequently in P2P energy trading with Prosumer 1, which reduces their costs by purchasing electricity at night and selling excess electricity at noon. To accommodate excess PV generation at noon, Prosumer 9 needs to rent a larger capacity of shared ESS, charge the ESS when PV generation exceeds demand, and discharge the ESS when electricity purchase prices are high. 

This also shows the effect of the proposed mechanism. For prosumers with relatively scarce resources, such as wind power in the case study, they can engage in the P2P energy trading to earn profits and rent less shared ESS. For others, more rental shared ESS is needed to adjust the surplus and homogeneous generation.

\subsection{Convergence and privacy protection of proposed algorithm}

\begin{table}[!t]
    \centering
    \caption{Comparison of different algorithms}
    \label{Algorithm Efficiency}
    \begin{tabular}{l c c c}
        \toprule
           Algorithm & Total time & Parallel time & \makecell{Privacy \\ protection}\\
        \midrule
           This paper & 420s &  89s & \ding{51}\\
           ADMM with (\ref{two-block})  & 434s & 87s & \ding{55}\\
           ADMM with NI-function & 40min & 7min & \ding{55}\\
        \bottomrule
    \end{tabular}
    
\end{table}

Table. \ref{Algorithm Efficiency} compares different solution algorithms of the GNE. The ADMM based on NI-function is a distributed algorithm to solve the GNE of the problem (\ref{Initial problem}), which is similar to the algorithm used in \cite{xiao2022new}. Instead of transforming the generalized Nash game into a QP problem, the algorithm applies NI-function to solve the equilibrium directly, which is applicable to the solution of variational equilibrium in the vast majority of generalized Nash games. However, this method is considerably slower due to its two-layer iterative loop structure and lacks effective privacy protection. The algorithm proposed in this paper, while slightly slower than the standard ADMM, offers enhanced privacy protection. Notably, the sub-problems for prosumers can be computed in parallel, as can the sub-problems for the shared ESS and the P2P transaction center. By using the maximum iteration time required by prosumers as the parallel computation time, the process is optimized. The day-ahead problem can be resolved in just 89 seconds, which is deemed acceptable.

\begin{figure}[!t]
    \centering
    \includegraphics{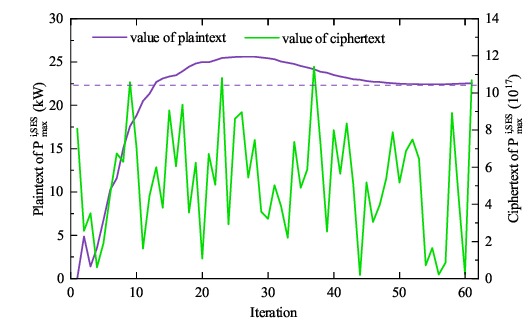}
    \caption{The value of plaintext and ciphertext in each iteration. The dashed line is the variational equilibrium obtained by centralized solution.}
    \label{Pailliertext}
\end{figure}
    
\begin{figure}
    \centering
    \includegraphics[trim = 2 2 2 2,clip]{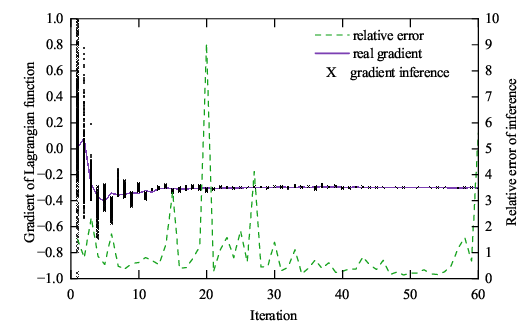}
    \caption{Estimated gradient of Lagrangian function and relative error of inference of 2000 trials. The relative error is the ratio of average error and the absolute value of the difference between the gradient of the Lagrangian function in the iteration and the gradient of the Lagrangian function in the equilibrium state}
    \label{NablaInfer}
\end{figure}

Fig. \ref{Pailliertext} compares the plaintext and the ciphertext of $P_{max}^{2,SES}(k)$ to show the convergence and effectiveness of privacy protection. It shows that the variable converges in 62 iterations. Prosumer 2 transmits the data to the shared ESS in ciphertext form. Unlike the regular sequence of plaintext data, the ciphertext appears irregular and does not readily reveal any information. 

Shared ESS can generate estimates of $\tau_P^i(k)$, and then obtain a series of estimated value of variables (e.g. $P^{i,SES}_{max}(k)$) and value of gradient of Lagrangian function according to (\ref{Encry prosumer update}) (\ref{alpha compute}) (\ref{Encry dual update}).

Fig. \ref{NablaInfer} shows the shared ESS's inference of gradient of the Lagrangian function of prosumer 2: $\nabla_{P_{max}^{2,SES}} L_0^2(\bm{x}^2)= \nabla_{P_{max}^{2,SES}} C_0^2(\bm{x}^2) + {\bm{\lambda}^2}^T\nabla_{P_{max}^{2,SES}}\bm{g}(\bm{x^2})$. The relative error is defined as the ratio of average error and the absolute value of the difference between the gradient of the Lagrangian function in the iteration and the gradient of the Lagrangian function in the equilibrium state. The average relative error is about 100\%, and in most iterations the relative error is greater than 20\%. This means that the shared ESS cannot infer private data of prosumer 2, and highlights the effectiveness of the privacy protection.

\section{Conclusion}
\label{Conclusion section}

The paper proposes a capacity renting framework for shared energy storage systems considering the interactions of prosumers. In the framework, The prosumers are directly coupled by the energy transactions, but are decoupled in the communication network for better distributed optimization. The generalized Nash game based model can effectively describe the competitive relationship among prosumers.
Based on the results obtained, several conclusions can be drawn:

(1) Integrating peer-to-peer energy trading into the capacity renting framework of shared energy storage systems can further increase social welfare. In the case study, the increase in value due to P2P energy trading is approximately 50\% of the value provided by the shared ESS alone. Although P2P energy trading slightly reduces the revenue of the shared ESS, it benefits prosumers and improves the consistency of charge-discharge behavior, leading to greater efficiency of the shared ESS.

(2) The distributed solution algorithm effectively determines the generalized Nash equilibrium (GNE) as fast as the traditional ADMM while maintaining participants' privacy. By converting the generalized Nash game into a quadratic QP problem and solving it with the ADMM algorithm integrated with Paillier Cryptosystem, the day-ahead problem can be solved in under 2 minutes, ensuring privacy protection. Despite potential communication data vulnerabilities, the encrypted data prevents the inference of private information.

Future research may further improve the proposed capacity renting mechanism with following aspects. 1) Pricing strategies of shared ESS. Shared ESS can set price to earn more profit by methods such as Stakelberg game. 2) Consideration of other types of energy storage. In this research, only batteries are considered and shared. The sharing mechanism appliable on other types of ESS is required.

\section*{Acknowledgement}
This work is supported by the National Natural Science Foundation of China Regional Joint Program (U22A20224).

\bibliographystyle{elsarticle-num} 
\bibliography{reference.bib}

\begin{thebibliography}{10}
\expandafter\ifx\csname url\endcsname\relax
  \def\url#1{\texttt{#1}}\fi
\expandafter\ifx\csname urlprefix\endcsname\relax\def\urlprefix{URL }\fi
\expandafter\ifx\csname href\endcsname\relax
  \def\href#1#2{#2} \def\path#1{#1}\fi

\bibitem{AI2022119968}
W.~Ai, T.~Deng, W.~Qi, Farsighted stability of distributed energy resource sharing, Applied Energy 326 (2022) 119968.

\bibitem{grvzanic2022prosumers}
M.~Gr{\v{z}}ani{\'c}, T.~Capuder, N.~Zhang, W.~Huang, Prosumers as active market participants: A systematic review of evolution of opportunities, models and challenges, Renewable and Sustainable Energy Reviews 154 (2022) 111859.

\bibitem{choudhury2022review}
S.~Choudhury, Review of energy storage system technologies integration to microgrid: Types, control strategies, issues, and future prospects, Journal of Energy Storage 48 (2022) 103966.

\bibitem{dai2021utilization}
R.~Dai, R.~Esmaeilbeigi, H.~Charkhgard, The utilization of shared energy storage in energy systems: A comprehensive review, IEEE Transactions on Smart Grid 12~(4) (2021) 3163--3174.

\bibitem{wang2024two}
Z.~Wang, L.~Chen, X.~Li, S.~Mei, A two-stage optimization approach-based energy storage sharing strategy selection for limited rational users, Journal of Energy Storage 93 (2024) 112098.

\bibitem{xiao2022new}
J.~Xiao, Y.~Yang, S.~Cui, X.~Liu, A new energy storage sharing framework with regard to both storage capacity and power capacity, Applied Energy 307 (2022) 118171.

\bibitem{ma2024optimal}
L.~Ma, X.~Li, X.~Kong, C.~Yang, L.~Chen, Optimal participation and cost allocation of shared energy storage considering customer directrix load demand response, Journal of Energy Storage 81 (2024) 110404.

\bibitem{zhao2019virtual}
D.~Zhao, H.~Wang, J.~Huang, X.~Lin, Virtual energy storage sharing and capacity allocation, IEEE transactions on smart grid 11~(2) (2019) 1112--1123.

\bibitem{chen2022cooperative}
C.~Chen, Y.~Li, W.~Qiu, C.~Liu, Q.~Zhang, Z.~Li, Z.~Lin, L.~Yang, Cooperative-game-based day-ahead scheduling of local integrated energy systems with shared energy storage, IEEE Transactions on Sustainable Energy 13~(4) (2022) 1994--2011.

\bibitem{jo2020demand}
J.~Jo, J.~Park, Demand-side management with shared energy storage system in smart grid, IEEE transactions on smart grid 11~(5) (2020) 4466--4476.

\bibitem{lai2021individualized}
S.~Lai, J.~Qiu, Y.~Tao, Individualized pricing of energy storage sharing based on discount sensitivity, IEEE Transactions on Industrial Informatics 18~(7) (2021) 4642--4653.

\bibitem{zheng2022peer}
B.~Zheng, W.~Wei, Y.~Chen, Q.~Wu, S.~Mei, A peer-to-peer energy trading market embedded with residential shared energy storage units, Applied Energy 308 (2022) 118400.

\bibitem{SUN2024110406}
B.~Sun, R.~Jing, Y.~Zeng, W.~Wei, X.~Jin, B.~Huang, Three-side coordinated dispatching method for intelligent distribution network considering dynamic capacity division of shared energy storage system, Journal of Energy Storage 81 (2024) 110406.

\bibitem{LI2022104710}
L.~Li, X.~Cao, S.~Zhang, Shared energy storage system for prosumers in a community: Investment decision, economic operation, and benefits allocation under a cost-effective way, Journal of Energy Storage 50 (2022) 104710.

\bibitem{9316966}
Q.~Li, J.~S. Gundersen, R.~Heusdens, M.~G. Christensen, Privacy-preserving distributed processing: Metrics, bounds and algorithms, IEEE Transactions on Information Forensics and Security 16 (2021) 2090--2103.

\bibitem{zhou2023Coordination}
X.~Zhou, H.~He, B.~Wang, X.~Zhu, H.~Sun, Coordination-decomposition algorithm based on paillier encryption and consensus mechanism (in chinese), Journal of Global Energy Interconnection 6 (2023) 362--369.

\bibitem{pena2022integration}
A.~Pena-Bello, D.~Parra, M.~Herberz, V.~Tiefenbeck, M.~K. Patel, U.~J. Hahnel, Integration of prosumer peer-to-peer trading decisions into energy community modelling, Nature Energy 7~(1) (2022) 74--82.

\bibitem{9618645}
Y.~Chen, W.~Wei, M.~Li, L.~Chen, J.~P.~S. Catalão, Flexibility requirement when tracking renewable power fluctuation with peer-to-peer energy sharing, IEEE Transactions on Smart Grid 13~(2) (2022) 1113--1125.

\bibitem{chen2023region}
Y.~Chen, N.~Liu, L.~Chen, X.~Yu, Region-to-region energy sharing for prosumer clusters in distribution network: A multi-leaders and multi-followers stackelberg game, IEEE Transactions on Energy Markets, Policy and Regulation (2023).

\bibitem{yan2023distributed}
D.~Yan, Y.~Chen, Distributed coordination of charging stations with shared energy storage in a distribution network, IEEE Transactions on Smart Grid (2023).

\bibitem{7525222}
E.~Nozari, P.~Tallapragada, J.~Cortés, Differentially private distributed convex optimization via objective perturbation, in: 2016 American Control Conference (ACC), 2016, pp. 2061--2066.

\bibitem{LU2018314}
Y.~Lu, M.~Zhu, Privacy preserving distributed optimization using homomorphic encryption, Automatica 96 (2018) 314--325.

\bibitem{10589503}
Q.~Xie, S.~Jiang, L.~Jiang, Y.~Huang, Z.~Zhao, S.~Khan, W.~Dai, Z.~Liu, K.~Wu, Efficiency optimization techniques in privacy-preserving federated learning with homomorphic encryption: A brief survey, IEEE Internet of Things Journal 11~(14) (2024) 24569--24580.

\bibitem{zhang2018admm}
C.~Zhang, M.~Ahmad, Y.~Wang, Admm based privacy-preserving decentralized optimization, IEEE Transactions on Information Forensics and Security 14~(3) (2018) 565--580.

\bibitem{8864035}
Y.~Chen, S.~Mei, F.~Zhou, S.~H. Low, W.~Wei, F.~Liu, An energy sharing game with generalized demand bidding: Model and properties, IEEE Transactions on Smart Grid 11~(3) (2020) 2055--2066.

\bibitem{he2021new}
X.~He, J.-W. Xiao, S.-C. Cui, X.-K. Liu, Y.-W. Wang, A new cooperation framework with a fair clearing scheme for energy storage sharing, IEEE Transactions on Industrial Informatics 18~(9) (2021) 5893--5904.

\bibitem{ullah2021peer}
M.~H. Ullah, J.-D. Park, Peer-to-peer energy trading in transactive markets considering physical network constraints, IEEE Transactions on Smart Grid 12~(4) (2021) 3390--3403.

\bibitem{paillier1999public}
P.~Paillier, Public-key cryptosystems based on composite degree residuosity classes, in: International conference on the theory and applications of cryptographic techniques, Springer, 1999, pp. 223--238.

\bibitem{deng2017parallel}
W.~Deng, M.-J. Lai, Z.~Peng, W.~Yin, Parallel multi-block admm with o (1/k) convergence, Journal of Scientific Computing 71 (2017) 712--736.

\bibitem{ShandongPrice}
{State grid Shandong electric power company}, {Announcement of Industrial and Commercial time-of-use Tariff for 2024 (in Chinese)}, https://www.pvmeng.com/2023/12/07/22402/ (2024).

\bibitem{facchinei2010generalized}
F.~Facchinei, C.~Kanzow, Generalized nash equilibrium problems, Annals of Operations Research 175~(1) (2010) 177--211.

\end{thebibliography}

\section*{Appendix A}

\setcounter{equation}{0}
\renewcommand{\theequation}{A.\arabic{equation}}

In the model of the generalized Nash game, it can be found that $\bm{X^i_0}, \bm{X^i_1}$ are both affine spaces. Therefore, the problem (\ref{Initial problem}) can be abstracted as:
\begin{align}
    \min_{\bm{x^i}} \quad & C^{i}(\bm{x^i}, \bm{x^{-i}}) = C^{i}_0(\bm{x^i_0}) + C^{i}_1(\bm{x^i_1}, \bm{x^{-i}_1})\notag \\
    s.t. \quad & \bm{A_0^i} \bm{x_0^i} + 
    \begin{bmatrix}
        \bm{A_p^i} & \bm{A_s^i}
    \end{bmatrix}
    \begin{bmatrix}
        P_{max}^{i,SES} \\\\
        S_{max}^{i,SES}
    \end{bmatrix}
    \leq \bm{b^i} : \bm{\mu^i_0} \notag \\
     \sum_{j \in \mathbb{S}} & ( \bm{E_0^j} \bm{x_0^j} + 
    \begin{bmatrix}
        \bm{E_p^j} & \bm{E_s^j}
    \end{bmatrix}
    \begin{bmatrix}
        P_{max}^{j,SES} \\\\
        S_{max}^{j,SES}
    \end{bmatrix})
    \leq \bm{f} : \bm{\mu^i_1}
    \label{abstracted problem}
\end{align}
where $\bm{A_0^i}, \bm{A_p^i}, \bm{A_s^i}, \bm{E_0^j}, \bm{E_p^j}, \bm{E_s^j}$ are coefficient matrixes,  $\bm{b^i}, \bm{f}$ are coefficient vectors, and $\bm{\mu^i_0}, \bm{\mu^i_1}$ are dual variables of the constraints.

The KKT system of problem (\ref{abstracted problem}) can be written as:
\begin{align}
    &\nabla_{\bm{x^i_0}} C^{i}_0(\bm{x^i_0}) + {\bm{A_0^i}}^T \bm{\mu^i_0} + {\bm{E_0^i}}^T \bm{\mu^i_1} = 0\notag \\
    &a_p + b_p\sum_{j \in \mathbb{S}}P^{j,SES}_{max} + b_p P^{i,SES}_{max}  + {\bm{A_p^i}}^T \bm{\mu^i_0} + {\bm{E_p^i}}^T \bm{\mu^i_1} = 0 \notag \\
    &a_s + b_s\sum_{j \in \mathbb{S}}S^{j,SES}_{max} + b_s S^{i,SES}_{max}  + {\bm{A_s^i}}^T \bm{\mu^i_0} + {\bm{E_s^i}}^T \bm{\mu^i_1} = 0 \notag \\
    & 0 \leq \bm{\mu_0^i} \perp (\bm{A_0^i} \bm{x_0^i} + 
    \begin{bmatrix}
        \bm{A_p^i} & \bm{A_s^i}
    \end{bmatrix}
    \begin{bmatrix}
        P_{max}^{i,SES} \\\\
        S_{max}^{i,SES}
    \end{bmatrix}
    - \bm{b^i}) \leq 0 \notag \\
    &0 \leq \bm{\mu_1^i} \perp (\sum_{j \in \mathbb{S}}( \bm{E_0^j} \bm{x_0^j} + 
    \begin{bmatrix}
        \bm{E_p^j} & \bm{E_s^j}
    \end{bmatrix}
    \begin{bmatrix}
        P_{max}^{j,SES} \\\\
        S_{max}^{j,SES}
    \end{bmatrix})
    - \bm{f}) \leq 0 \notag \\
    \label{KKTsystems}
\end{align}
where $0 \leq \bm{a} \perp \bm{b} \leq 0$ means $\bm{a}^T\bm{b} = 0, \bm{a} \geq \bm{0}, \bm{b} \leq \bm{0}$.

To solve the variational equilibrium, the dual variables for the common constraints of each prosumer should satisfy:
\begin{equation}
    \bm{\mu_1^i} = \bm{\mu_1}
    \label{variational KKT}
\end{equation}

It can be easily proved that the objective of the generalized Nash game (\ref{abstracted problem}) $C^i(\bm{x^i},\bm{x^{-i}})$ is convex over $\bm{x^i}$. The constraints are affine, which satisfies Slater condition. Therefore, the variational equilibrium of (\ref{abstracted problem}) is equivalent to the solution of the KKT system (\ref{KKTsystems}), (\ref{variational KKT}) \cite{facchinei2010generalized}.

It can be noticed that the problem (\ref{Centralized abstract problem}) has the same KKT system with (\ref{KKTsystems}) (\ref{variational KKT})
\begin{align}
    \min_{\bm{x}} & \quad \sum_{i \in \mathbb{S}} C_0^i(\bm{x_0^i}) + \frac{1}{2} b_s (\sum_{i \in \mathbb{S}} S_{max}^{i,SES})^2 + a_s(\sum_{i \in \mathbb{S}} S_{max}^{i,SES}) \notag \\
    &+ \frac{1}{2}b_s(\sum_{i \in \mathbb{S}} {S_{max}^{i,SES}}^2)  + \frac{1}{2} b_p (\sum_{i \in \mathbb{S}} P_{max}^{i,SES})^2 \notag \\
    &+ a_p(\sum_{i \in \mathbb{S}} P_{max}^{i,SES}) + \frac{1}{2}b_p(\sum_{i \in \mathbb{S}} {P_{max}^{i,SES}}^2) \notag \\
    s.t. & \quad \bm{A_0^i} \bm{x_0^i} + 
    \begin{bmatrix}
        \bm{A_p^i} & \bm{A_s^i}
    \end{bmatrix}
    \begin{bmatrix}
        P_{max}^{i,SES} \\\\
        S_{max}^{i,SES}
    \end{bmatrix}
    \leq \bm{b^i} \quad \forall i \in \mathbb{S} \notag \\
    & \sum_{i \in \mathbb{S}}  ( \bm{E_0^i} \bm{x_0^i} + 
    \begin{bmatrix}
        \bm{E_p^i} & \bm{E_s^i}
    \end{bmatrix}
    \begin{bmatrix}
        P_{max}^{i,SES} \\\\
        S_{max}^{i,SES}
    \end{bmatrix})
    \leq \bm{f}
    \label{Centralized abstract problem}
\end{align}

Problem (\ref{Centralized abstract problem}) is a QP problem, and therefore the optimal solution of problem (\ref{Centralized abstract problem}) is equivalent to the KKT system. Therefore, the variational equilibrium of (\ref{Initial problem}) is equivalent to the optimal solution of problem (\ref{Centralized abstract problem}), which is the same as problem (\ref{Centralized problem}). It can be deduced that the variational equilibrium of (\ref{Initial problem}) exists if and only if problem (\ref{Centralized problem}) has a solution, and the equilibrium is the same as the optimal solution of problem (\ref{Centralized problem}).

\section*{Appendix B}

\setcounter{equation}{0}
\renewcommand{\theequation}{B.\arabic{equation}}

\setcounter{algocf}{0}
\renewcommand{\thealgocf}{B\arabic{algocf}}

The algorithm for standard ADMM is shown in Algorithm \ref{Standard ADMM}. For each prosumer, they need to update their decision variables by problem (\ref{Standard prosumer update}).

\begin{align}
    \bm{x^{i}}&(k+1) = \arg \min_{\bm{x^i} \in \bm{X_0^i}} \quad C^{i}_0(\bm{x^i}) \notag \\
    &+ \frac{\rho_P^i}{2} \parallel\tilde{P}^{i,SES}_{max}(k) - {P}^{i,SES}_{max} + \frac{{\mu}_P^i(k)}{\rho_P^i}\parallel_2^2 \notag \\
    &+ \frac{\rho_S^i}{2} \parallel\tilde{S}^{i,SES}_{max}(k) - {S}^{i,SES}_{max} + \frac{{\mu}_S^i(k)}{\rho_S^i}\parallel_2^2 \notag \\
    &+ \sum_{t}\sum_{j \in \mathbb{S}}\frac{\rho_{P2P}^i}{2} \parallel\tilde{P}^{i,j}_{t}(k) - {P}^{i,j}_{t} + \frac{{\mu}^{i,j}_t(k)}{\rho_{P2P}^i}\parallel_2^2
    \label{Standard prosumer update}
\end{align}
where ${\mu}_P^i, {\mu}_S^i$ and $ {\mu}^{i,j}_t$ are the dual variables of (\ref{SES-P assi equality}),(\ref{SES-S assi equality}), and (\ref{P2P assi equality}), respectively. $\rho_P^i, \rho_S^i, \rho_{P2P}^i$ are constant coefficients.

Then, the prosumers transmit ${P}^{i,SES}_{max}(k+1), {S}^{i,SES}_{max}(k+1)$ to the shared ESS, and transmit $\bm{P_{P2P}^i}(k+1)$ to the P2P transaction center. The shared ESS updates the assistant variables by problem (\ref{Standard SES-P update}) and (\ref{Standard SES-S update}).

\begin{align}
    \bm{\tilde{P}^{SES}}&(k+1) = \arg \min_{\bm{\tilde{P}^{SES}}} \frac{1}{2} b_p (\sum_{i \in \mathbb{S}} \tilde{P}_{max}^{i,SES})^2 \notag \\
    &+ a_p(\sum_{i \in \mathbb{S}} \tilde{P}_{max}^{i,SES}) + \frac{1}{2}b_p(\sum_{i \in \mathbb{S}} \tilde{P}{{}_{max}^{i,SES}}^2) \notag \\
    &+ \sum_{i \in \mathbb{S}} \frac{\rho_P^i}{2} \parallel\tilde{P}^{i,SES}_{max} - {P}^{i,SES}_{max}(k+1) + \frac{{\mu}_P^i(k)}{\rho_P^i}\parallel_2^2 \notag \\
    &s.t. \quad eqs. (\ref{SES-P constrain})
    \label{Standard SES-P update}
\end{align}
\begin{align}
    \bm{\tilde{S}^{SES}}&(k+1) = \arg \min_{\bm{\tilde{S}^{SES}}} \frac{1}{2} b_s (\sum_{i \in \mathbb{S}} \tilde{S}_{max}^{i,SES})^2 \notag \\
    &+ a_s(\sum_{i \in \mathbb{S}} \tilde{S}_{max}^{i,SES}) + \frac{1}{2}b_s(\sum_{i \in \mathbb{S}} \tilde{S}{{}_{max}^{i,SES}}^2)  \notag \\
    &+ \sum_{i \in \mathbb{S}} \frac{\rho_S^i}{2} \parallel\tilde{S}^{i,SES}_{max} - {S}^{i,SES}_{max}(k+1) + \frac{{\mu}_S^i(k)}{\rho_S^i}\parallel_2^2  \notag \\
    & s.t. \quad eqs. (\ref{SES-S constrain})
    \label{Standard SES-S update}
\end{align}
where $\bm{\tilde{P}^{SES}} = [\tilde{P}_{max}^{i,SES} \quad \forall i \in \mathbb{S}]$, $\bm{\tilde{S}^{SES}} = [\tilde{S}_{max}^{i,SES} \quad \forall i \in \mathbb{S}]$

The P2P transaction center updates the assistant variables by problem (\ref{Standard P2P center update}).
\begin{align}
    \bm{\tilde{P}_{P2P}}&(k+1) = \arg \min_{\bm{\tilde{P}_{P2P}}} \notag \\
    &\sum_{i,j \in \mathbb{S}}\frac{\rho_{P2P}^i}{2} \parallel\tilde{P}^{i,j}_{t} - {P}^{i,j}_{t}(k+1) + \frac{{\mu}^{i,j}_t(k)}{\rho_{P2P}^i}\parallel_2^2 \notag \\
    & s.t. \quad eqs. (\ref{P2P center constrains})
    \label{Standard P2P center update}
\end{align}
where $\bm{\tilde{P}_{P2P}} = [\tilde{P}^{i,j}_{t} \quad \forall i,j \in \mathbb{S}]$

Then, the dual variables are updated by eqs. (\ref{Standard dual update}).
\begin{align}
    &\mu_P^{i}(k+1) = {\mu}_P^i(k) + \rho_P^i (\tilde{P}^{i,SES}_{max}(k+1) - {P}^{i,SES}_{max}(k+1)) \notag \\
    &\mu_S^{i}(k+1) = {\mu}_S^i(k) + \rho_S^i (\tilde{S}^{i,SES}_{max}(k+1) - {S}^{i,SES}_{max}(k+1)) \notag \\
    &\mu^{i,j}_t(k+1) = {\mu}^{i,j}_t(k) + \rho_{P2P}^i (\tilde{P}^{i,j}_{t}(k+1) - {P}^{i,j}_{t}(k+1)) \notag \\
    \label{Standard dual update}
\end{align}

The gap can be calculated by:
\begin{align}
    c(k+1) &= \sum_{i \in \mathbb{S}}\frac{1}{\rho_P^i}\parallel\mu_P^{i}(k+1) - \mu_P^{i}(k)\parallel_2^2 \notag \\
    &+ \sum_{i \in \mathbb{S}}\frac{1}{\rho_S^i}\parallel\mu_S^{i}(k+1) - \mu_S^{i}(k)\parallel_2^2 \notag \\
    &+ \sum_{i,j \in \mathbb{S}}\frac{1}{\rho_{P2P}^i}\parallel\mu_t^{i,j}(k+1) - \mu_t^{i,j}(k)\parallel_2^2 \notag \\
    &+ \sum_{i \in \mathbb{S}}\rho_P^i \parallel\tilde{P}^{i,SES}_{max}(k+1) - \tilde{P}^{i,SES}_{max}(k)\parallel_2^2 \notag \\
    &+ \sum_{i \in \mathbb{S}}\rho_S^i \parallel\tilde{S}^{i,SES}_{max}(k+1) - \tilde{S}^{i,SES}_{max}(k)\parallel_2^2 \notag \\
    &+ \sum_{i,j \in \mathbb{S}}\rho_{P2P}^i \parallel\tilde{P}^{i,j}_{t}(k+1) - \tilde{P}^{i,j}_{t}(k)\parallel_2^2 \notag \\
    \label{Standard gap calculation}
\end{align}

For prosumers, they need to transmit ${P}^{i,SES}_{max}(k), {S}^{i,SES}_{max}(k)$ to the shared ESS, who can infer the value of the gradient of the Lagrangian function of prosumers from the sequential data. According to (\ref{Standard prosumer update}), equation (\ref{Privacy Infer}) is always satisfied. It shows that shared ESS can infer the gradient of the Lagrangian function of prosumers. According to a series of variable values and the corresponding gradient values, the function can be inferred, which means the privacy of prosumers is disclosed to the shared ESS.

\begin{align}
    &\nabla_{P^{i,SES}_{max}(k+1)}\quad L_0^i(\bm{x^i}) \notag \\
    &= -\mu_P^i(k) - \rho_P^i(\tilde{P}^{i,SES}_{max}(k) - {P}^{i,SES}_{max}(k+1))
    \label{Privacy Infer}
\end{align}
where $L_0^i(\bm{x^i})$ is the Lagrangian function of prosumer $i$.

\begin{algorithm}
Initialize: 
$\tilde{P}^{i,SES}_{max}(1)=0$,
$\tilde{S}^{i,SES}_{max}(1)=0$,
$\bm{\tilde{P}_{P2P}} = \bm{0}$,
${\mu}_P^i(1)=0$, ${\mu}_S^i(1)=0$, ${\mu}^{i,j}_t(1)=0$, $\rho > 0$, $c(1) = 10000$, $\epsilon$

\For{$k=1,2,3...$}{

\For{prosumer $i \in \mathbb{S}$}{
    update $\bm{x^i}(k+1)$ by problem \ref{Standard prosumer update}.
}
update assistant variables by problem \ref{Standard SES-P update}, \ref{Standard SES-S update} and \ref{Standard P2P center update}.

update dual variables by eqs. \ref{Standard dual update}.

calculate $c(k+1)$ by eqs. \ref{Standard gap calculation}

\If{$c(k+1) < \epsilon$}{
break
}

}

\caption{Algorithm of standard ADMM}
\label{Standard ADMM}
\end{algorithm}

\section*{Appendix C}

\setcounter{equation}{0}
\renewcommand{\theequation}{C.\arabic{equation}}
\setcounter{algocf}{0}
\renewcommand{\thealgocf}{C\arabic{algocf}}

Take the update process of prosumers (\ref{Standard prosumer update}) as an example. First, a proximal term is added to transform the update process (\ref{Standard prosumer update}) into (\ref{Proximal prosumer update}). The convergence is proved in \cite{deng2017parallel}.
\begin{align}
    \bm{x^{i}}&(k+1) = \arg \min_{\bm{x^i}} \quad C^{i}_0(\bm{x^i}) \notag \\
    &+ \frac{\rho_P^i}{2} \parallel\tilde{P}^{i,SES}_{max}(k) - {P}^{i,SES}_{max} + \frac{{\mu}_P^i(k)}{\rho_P^i}\parallel_2^2 \notag \\
    &+ \frac{\rho_S^i}{2} \parallel\tilde{S}^{i,SES}_{max}(k) - {S}^{i,SES}_{max} + \frac{{\mu}_S^i(k)}{\rho_S^i}\parallel_2^2 \notag \\
    &+ \sum_{t}\sum_{j \in \mathbb{S}}\frac{\rho_{P2P}^i}{2} \parallel\tilde{P}^{i,j}_{t}(k) - {P}^{i,j}_{t} + \frac{{\mu}^{i,j}_t(k)}{\rho_{P2P}^i}\parallel_2^2 \notag \\
    &+ \frac{\gamma_P^i}{2}\parallel {P}^{i,SES}_{max} - {P}^{i,SES}_{max}(k)\parallel_2^2 \notag \\
    &+ \frac{\gamma_S^i}{2}\parallel {S}^{i,SES}_{max} - {S}^{i,SES}_{max}(k)\parallel_2^2 \notag \\
    &+\sum_{t}\sum_{j \in \mathbb{S}}\frac{\gamma_{P2P}^i}{2}\parallel {P}^{i,j}_{t} - {P}^{i,j}_{t}(k)\parallel_2^2 \notag \\
    s.t. & \quad \bm{g}(\bm{x^i}) \leq \bm{0}
    \label{Proximal prosumer update}
\end{align}
where the constraints $\bm{g}(\bm{x^i}) \leq \bm{0}$ are equivalent to the set $\bm{X_0^i}$. $\gamma_P^i, \gamma_S^i, \gamma_{P2P}^i \geq 0$ 

Apply KKT condition on problem (\ref{Proximal prosumer update}):
\begin{align}
    &\nabla_{P_{max}^{i,SES}} C_0^i(\bm{x}^i) + {\bm{\lambda}^i}^T\nabla_{P_{max}^{i,SES}}\bm{g}(\bm{x^i}) -  {\mu}_P^i(k) + \notag \\
    &\rho_P^i({P}^{i,SES}_{max} - \tilde{P}^{i,SES}_{max}(k) ) + \gamma_P^i({P}^{i,SES}_{max} - P^{i,SES}_{max}(k) ) = \notag \\
    &\nabla_{P_{max}^{i,SES}} C_0^i(\bm{x}^i) + {\bm{\lambda}^i}^T\nabla_{P_{max}^{i,SES}}\bm{g}(\bm{x^i}) + \alpha_P^i(k) -  {\mu}_P^i(k) + \notag \\
    &(\rho_P^i+\gamma_P^i)(P_{max}^{i,SES} - P_{max}^{i,SES}(k))= 0 \notag \\
    &\nabla_{S_{max}^{i,SES}} C_0^i(\bm{x}^i) + {\bm{\lambda}^i}^T\nabla_{S_{max}^{i,SES}}\bm{g}(\bm{x^i})+ \alpha_S^i(k) -  {\mu}_S^i(k) + \notag \\
    &(\rho_S^i+\gamma_S^i)(S_{max}^{i,SES} - S_{max}^{i,SES}(k))= 0 \notag \\
    &\nabla_{{P}^{i,j}_{t}} C_0^i(\bm{x}^i) + {\bm{\lambda}^i}^T\nabla_{{P}^{i,j}_{t}}\bm{g}(\bm{x^i}) +\alpha_t^{i,j}(k) -  {\mu}^{i,j}_t(k) + \notag \\
    &(\rho_{P2P}^i+\gamma_{P2P}^i)({P}^{i,j}_{t} - {P}^{i,j}_{t}(k)) 
    = 0 \quad \forall t, j \in \mathbb{S}  \notag \\
    &\nabla_{\bm{x_0^i}} C_0^i(\bm{x}^i) + {\bm{\lambda}^i}^T\nabla_{\bm{x_0^i}}\bm{g}(\bm{x^i}) = 0 \notag \\
    &\bm{g}(\bm{x^i}) \leq 0, \quad {\bm{\lambda}^i}^T\bm{g}(\bm{x^i}) = \bm{0}
    \label{KKT condition Proximal}
\end{align}
where ${\bm{\lambda}^i}$ is the dual variable of the constraints. $\beta_P^i, \beta_S^i, \beta_{P2P}^i, \alpha_P^i, \alpha_S^i, \alpha_{t}^{i,j}$ satisfy:
\begin{align}
    &\beta_P^i = \rho_P^i + \gamma_P^i  \notag \\
    &\beta_S^i = \rho_S^i + \gamma_S^i  \notag \\
    &\beta_{P2P}^i = \rho_{P2P}^i + \gamma_{P2P}^i  \notag \\
    &\alpha_P^i(k) = \rho_P^i(P_{max}^{i,SES}(k) - \tilde{P}_{max}^{i,SES}(k)) \notag \\
    &\alpha_S^i(k) = \rho_S^i(S_{max}^{i,SES}(k) - \tilde{S}_{max}^{i,SES}(k)) \notag \\
    &\alpha_{t}^{i,j}(k) = \rho_{P2P}^i(P_t^{i,j}(k) - \tilde{P}_t^{i,j}(k))
\end{align}

The KKT system of the optimization problem (\ref{Final prosumer update}) is also (\ref{KKT condition Proximal}). Therefore, the problem (\ref{Final prosumer update}) is equivalent to (\ref{Proximal prosumer update}).
\begin{align}
    &\bm{x^{i}}(k+1) = \arg \min_{\bm{x^i} \in \bm{X^i}} \quad C^{i}_0(\bm{x^i}) 
     \notag \\
    &+ \frac{\beta_P^i}{2}\parallel P_{max}^{i,SES} - P_{max}^{i,SES}(k)\parallel_2^2 + (\alpha_P^i(k)-\mu_P^i(k))P_{max}^{i,SES} \notag \\
    &+ \frac{\beta_S^i}{2}\parallel S_{max}^{i,SES} - S_{max}^{i,SES}(k)\parallel_2^2 + (\alpha_S^i(k)-\mu_S^i(k))S_{max}^{i,SES} \notag \\
    &+ \sum_{t}\sum_{j \in \mathbb{S}} (\frac{\beta_{P2P}^i}{2}\parallel {P}^{i,j}_{t} - {P}^{i,j}_{t}(k)\parallel_2^2 + (\alpha_t^{i,j}(k)-\mu_{t}^{i,j}(k)){P}^{i,j}_{t})
    \label{Final prosumer update}
\end{align}

Let $\rho_P^i, \rho_S^i, \rho_{P2P}^i$ be time-varying  \cite{zhang2018admm}. In each iteration, the coefficients are set at:
\begin{align}
    &\rho_P^i(k) = \tau_P^i(k)\tilde{\tau}_P^i(k) \notag \\
    &\rho_S^i(k) = \tau_S^i(k)\tilde{\tau}_S^i(k) \notag \\
    &\rho_{P2P}^i(k) = \tau_{P2P}^i(k)\tilde{\tau}_{P2P}^i(k)
\end{align}
where $\tau_P^i(k),\tilde{\tau}_P^i(k),\tau_S^i(k),\tilde{\tau}_S^i(k),\tau_{P2P}^i(k)\tilde,{\tau}_{P2P}^i(k)$ are random numbers. In this way, the value of $\rho_P^i, \rho_S^i, \rho_{P2P}^i$ are different in each iteration. Let $\beta_P^i, \beta_S^i, \beta_{P2P}^i$ be constant numbers so that $\gamma_P^i, \gamma_S^i, \gamma_{P2P}^i$ can be determined.

\section*{Appendix D}

\setcounter{equation}{0}
\renewcommand{\theequation}{D.\arabic{equation}}
\setcounter{algocf}{0}
\renewcommand{\thealgocf}{D\arabic{algocf}}

\begin{algorithm}[!t]
\textbf{Key generation}:

1) Choose two large prime numbers $p$ and $q$ of equal bit-length

2) Let $\Gamma = pq$, $\zeta = lcm(p-1,q-1)$ where $lcm$ means the least common multiple. Let $\Omega = n+1$.

3) Define $L(x) = \frac{x-1}{n}$. Let $\sigma = (L(\Omega^{\zeta} \; mod \;  n^2))^{-1}\; mod\; n$

4) The public key is $(\Gamma, \Omega)$, and the private key is $(\zeta, \sigma)$.

\textbf{Encryption} ($c = E(m), 0 \leq m \leq n$):

1) Choose a random number $r: 0 \leq r \leq n$.

2) $c = \Omega^m\cdot r^\Gamma \; mod \; \Gamma^2$

\textbf{Decryption} ($m = D(c)$):

1) $m = L(c^\zeta \; mod \; \Gamma^2) \cdot \sigma \; mod \; n$

\textbf{Homomorphic addition}

1) $E(m_1 + m_2) = E(m_1)\cdot E(m_2) \; mod \; \Gamma^2$

2) $E(am) = E(m)^a \; mod \; \Gamma^2$

\caption{Paillier Cryptosystem}
\label{Paillier Cryptosystem}
\end{algorithm}

Algorithm \ref{Paillier Cryptosystem} shows the Paillier cryptosystem, including the process of key generation, encryption, decryption, and the characteristics of homomorphic addition. 

Notes: the data to be encrypted must be a natural number. For negative integers, it should be encrypted in the form of 2's complement. 

\end{document}